\documentclass[fleqn,usenatbib]{mnras}

\usepackage{newtxtext,newtxmath}
\usepackage[T1]{fontenc}

\DeclareRobustCommand{\VAN}[3]{#2}
\let\VANthebibliography\thebibliography
\def\thebibliography{\DeclareRobustCommand{\VAN}[3]{##3}\VANthebibliography}

\usepackage{graphicx}     
\usepackage{amsmath}      

\usepackage{amssymb}      
\usepackage{subcaption}   

\title{Microscopy assessment of two multiple-transient candidates}

\author[Yang \& Villarroel]{%
Jun Yang$^{1}$\thanks{E-mail: junyangpro@gmail.com} and Beatriz Villarroel$^{2}$\\
$^{1}$ Astronomy Department, Harvard University, 60 Garden St., Cambridge, MA 02138, USA\\
$^{2}$ Nordita, KTH Royal Institute of Technology and Stockholm University, Hannes Alfv\'ens v\"ag 12, SE-106 91 Stockholm, Sweden
}

\date{Accepted XXX. Received YYY; in original form ZZZ}
\pubyear{2026}

\begin{document}
\label{firstpage}
\pagerange{\pageref{firstpage}--\pageref{lastpage}}
\maketitle

\begin{abstract}
We present a microscopy-based assessment of two representative cases of multiple-transient candidates identified on historical photographic plates: the nine-transient event of 12 April 1950 (XE325) and the triple-transient event of 19 July 1952 (XE186). These events consist of spatially clustered, point-like sources detected within single long-exposure plates, lacking the streaks expected from moving objects and therefore suggestive of short-duration flashes.

We analyse microscopy images of the original glass-plate material together with an independent copy plate and archival metadata, and we register these images against the published survey digitizations so that individual catalogued sources can be located. In both cases the candidates appear slightly sharper and more compact than typical stellar point spread functions, qualitatively consistent with sub-second to few-second emission during exposures of $\sim 50$~min, and they are reproducible across different representations of the same field, which disfavours a digitization-artifact origin.

However, microscopy of individual point-like features cannot by itself separate a short-duration flash from an intrinsic emulsion defect, because the two scenarios produce qualitatively similar signatures at the single-source level. We also show that the multi-observatory parallax test, while powerful in principle, is not feasible for these two specific events with presently available material, owing to the absence of a simultaneous second-site plate reaching comparable depth. We conclude that resolving the nature of these events requires population-level statistics and, where future observations allow, purpose-built multi-station imaging. This work provides a proof-of-concept framework for combining microscopy with archival analysis in the study of historical transient phenomena.
\end{abstract}



\section{Introduction}

The identification of short-duration optical transients on historical photographic plates from Palomar Observatory has recently attracted renewed attention, particularly in the context of surveys searching for non-repeating or anomalous events. Early work on photographic plate archives demonstrated both the richness and the complexity of these datasets, where astronomers struggled to verify whether a point source was real or a coincidentally star-like plate defect, often suggesting the need for microscopy (see, e.g., \citealt{Greiner1990}).

Some years ago, \citet{Villarroel2021} reported the discovery of nine transient sources appearing simultaneously within a $\sim10'$ region on a Palomar Observatory Sky Survey (POSS-I) plate (XE325, 12 April 1950), while \citet{Solano2024} and subsequent work discussed a triple transient detected on 19 July 1952 (XE186). These events are characterised by point-like morphologies, lacking the streaks expected for moving objects during long exposures, suggesting very short-duration flashes on timescales of seconds or less. The morphologies in the digitised images are, on average, slightly sharper and rounder than those of normal stars on the plate (see \citealt{Hambly2024}). This difference falls within expectations for sub-second or second-long flashes from astronomical sources \citep{VillarroelARXIV}.

Most recently, \citet{Busko2026} provided evidence against plate-contamination hypotheses, such as plate defects and cosmic rays, by demonstrating that in telescopes exhibiting optical aberrations that distort stellar images, the transients exhibit the same distortions. Such behaviour is only expected if the light from the transient sources passes through the telescope optics.

Despite these intriguing properties, a fundamental ambiguity remains for individual point-like features: such features may arise either from genuine transient astrophysical (or near-Earth) phenomena, or from plate defects and emulsion irregularities. As emphasised in recent discussions \citep[e.g.][]{VillarroelPASP}, both detailed inspection of the original plates and statistical analyses of larger samples are needed to resolve this ambiguity. Statistical analyses make it possible to determine whether the transient population consists entirely of plate defects and cosmic rays, as alleged by \citet{Watters2026}, or whether it comprises a mixture of point-like contaminants and genuine transient sources.

Several statistical tests have been reported that are consistent with a non-zero fraction of genuine optical transients among point-like detections on photographic plates. These include improbable spatial clustering \citep{VillarroelPASP}, an apparent deficit of detections within the Earth's shadow \citep{VillarroelPASP,Doherty2026}, correlations with observing conditions (e.g. geomagnetic activity) \citep{Cann2026a,Cann2026b}, and the presence of morphologically similar candidates in independent plate archives \citep{Busko2026}. Interpreting such correlations remains an active topic of discussion in the literature, and motivates careful, plate-level verification of representative events.

In a very recent study, \citet{Busko2026b} demonstrated that on historical photographic plates obtained with a German telescope affected by severe optical aberrations, transient sources exhibited the same coma distortions as ordinary stellar images. The study also reported an example of a triple transient with coma distortions, as well as transients recorded on dates coinciding with nuclear tests. The presence of coma distortions in transients with morphologies and properties similar to those observed on the Palomar plates indicates that these features were produced by light passing through the telescope optics rather than by plate contamination. The discovery of a triple transient exhibiting coma distortions further supports the interpretation of the triple transient reported by \citet{Solano2024}.

In this work, we present a microscopy-based assessment of two representative cases: (1) the nine-transient event of 12 April 1950, and (2) the triple-transient event of 19 July 1952. Our aim is not to determine conclusively the nature of these events, but rather to investigate whether microscopic examination of the plate material provides discriminating evidence between transient flashes and plate defects, and to set out clearly what microscopy can and cannot establish.

\section{Methods}

Our analysis combines three complementary approaches: (i) direct microscopy inspection of plate material; (ii) comparison between original plates and copy plates; and (iii) temporal and positional matching using archival plate metadata, including registration of the microscopy fields against the published survey digitizations.

The plate identification was guided by archival searches using StarGlass and DASCH resources, combined with the published coordinates and dates of the transient candidates. The nine-transient case corresponds to 12 April 1950 (RA $\approx$ 14h11m, Dec $\approx +26^\circ$), while the triple-transient case corresponds to 19 July 1952 (RA $\approx$ 21h18m, Dec $\approx +50^\circ$). Matching plates from independent patrol observations were searched for by date and cross-checked against archival catalogues (Section~\ref{sec:crossmatch}).

To place the microscopy on a common footing with the discovery data, we registered each microscopy field-of-view to the corresponding published digitization --- the SuperCOSMOS and STScI scans presented by \citet{Villarroel2021} for XE325 and by \citet{Solano2024} for XE186 --- using the brighter field stars as fiducials. This registration allows each catalogued source to be located in the microscopy frame and inspected directly on the physical emulsion.
\label{sec:registration}

\subsection{Case 1: Inspection of original plate at Caltech}

For the nine-transient event (XE325), the original glass plate was analysed at Caltech, and images were obtained. The plate box was retrieved from storage by observatory staff and allowed to reach ambient temperature before opening. The plate has a cover glass carefully taped to it from when it was digitized approximately 30 years ago, which was left in place during the inspection. The plate corresponds to an exposure of approximately 50~min, during which nine point-like transient sources appear within a confined region.

\begin{figure}
    \centering
    \includegraphics[width=0.45\textwidth]{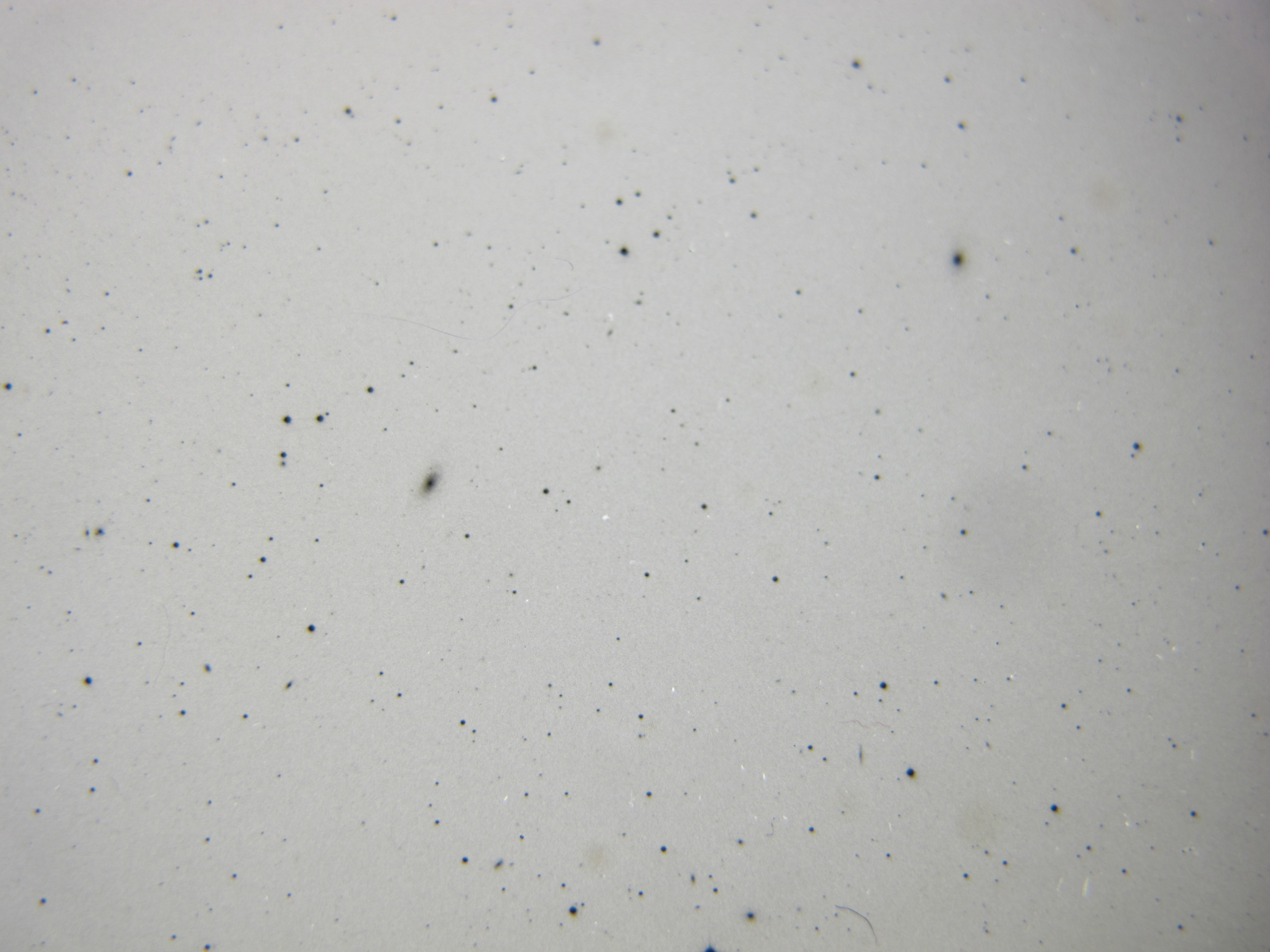}
    \caption{Microscopy image of the original photographic plate (XE325) containing the nine-transient event, taken at the Caltech plate archive. The image shows the field of interest with multiple point-like sources visible. Several compact features with sharp edges are present, including the transient candidates, contrasted against diffuse stellar images with characteristic halos. The field lies in the corner region of the plate, where emulsion defects are more common.}
    \label{fig:microscopy_9755}
\end{figure}

\begin{figure}
    \centering
    \includegraphics[width=0.45\textwidth]{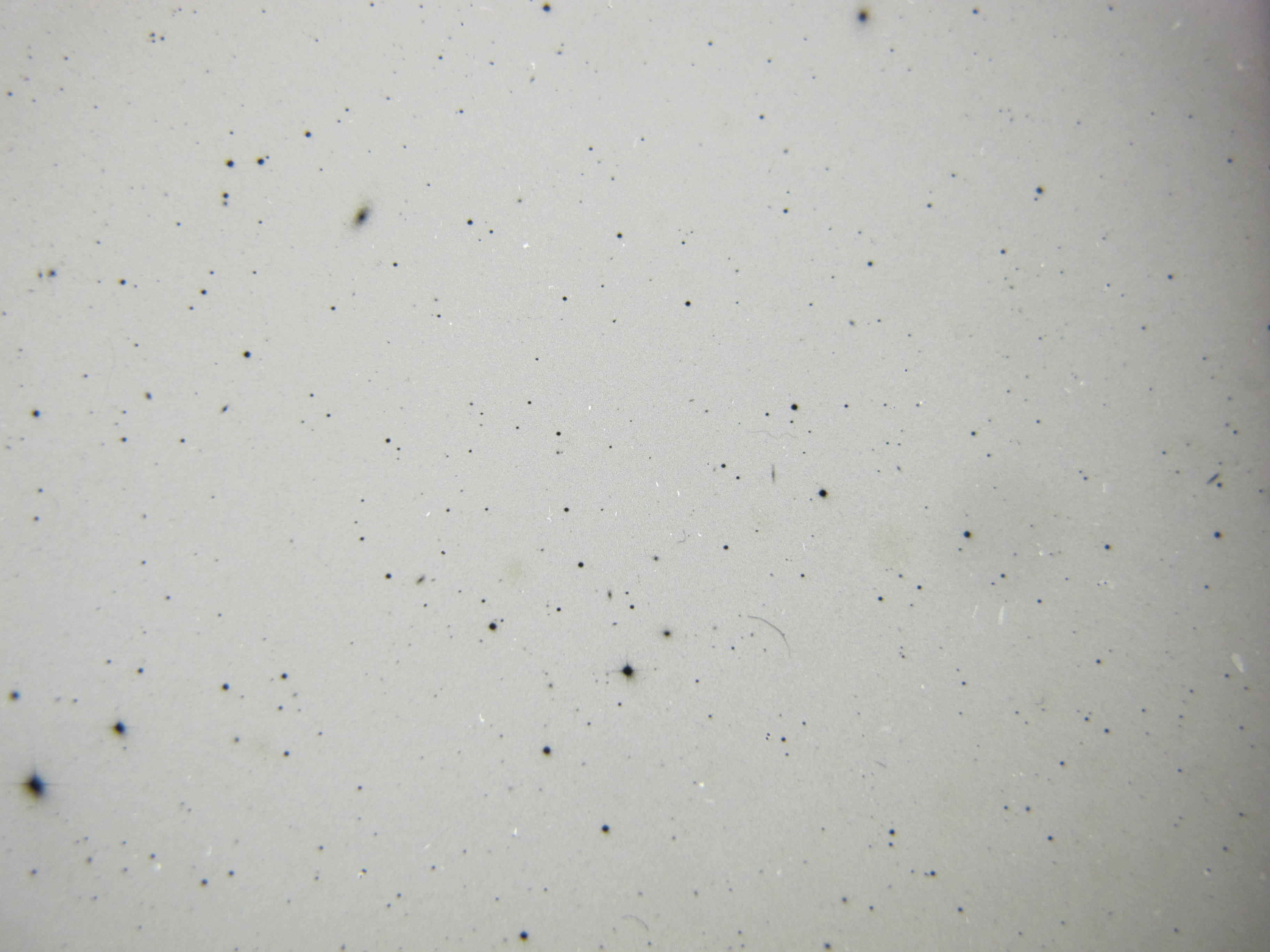}
    \caption{Second microscopy image of the XE325 plate field showing the distribution of sources. The transient candidates appear as sharp, round features without the elongation present in stellar images. The lack of elongation suggests an emission duration much shorter than the plate exposure time. Grain structure and potential emulsion irregularities are also visible, illustrating the challenge of distinguishing transient flashes from plate defects on morphology alone.}
    \label{fig:microscopy_9756}
\end{figure}

\begin{figure}
    \centering
    \includegraphics[width=0.45\textwidth]{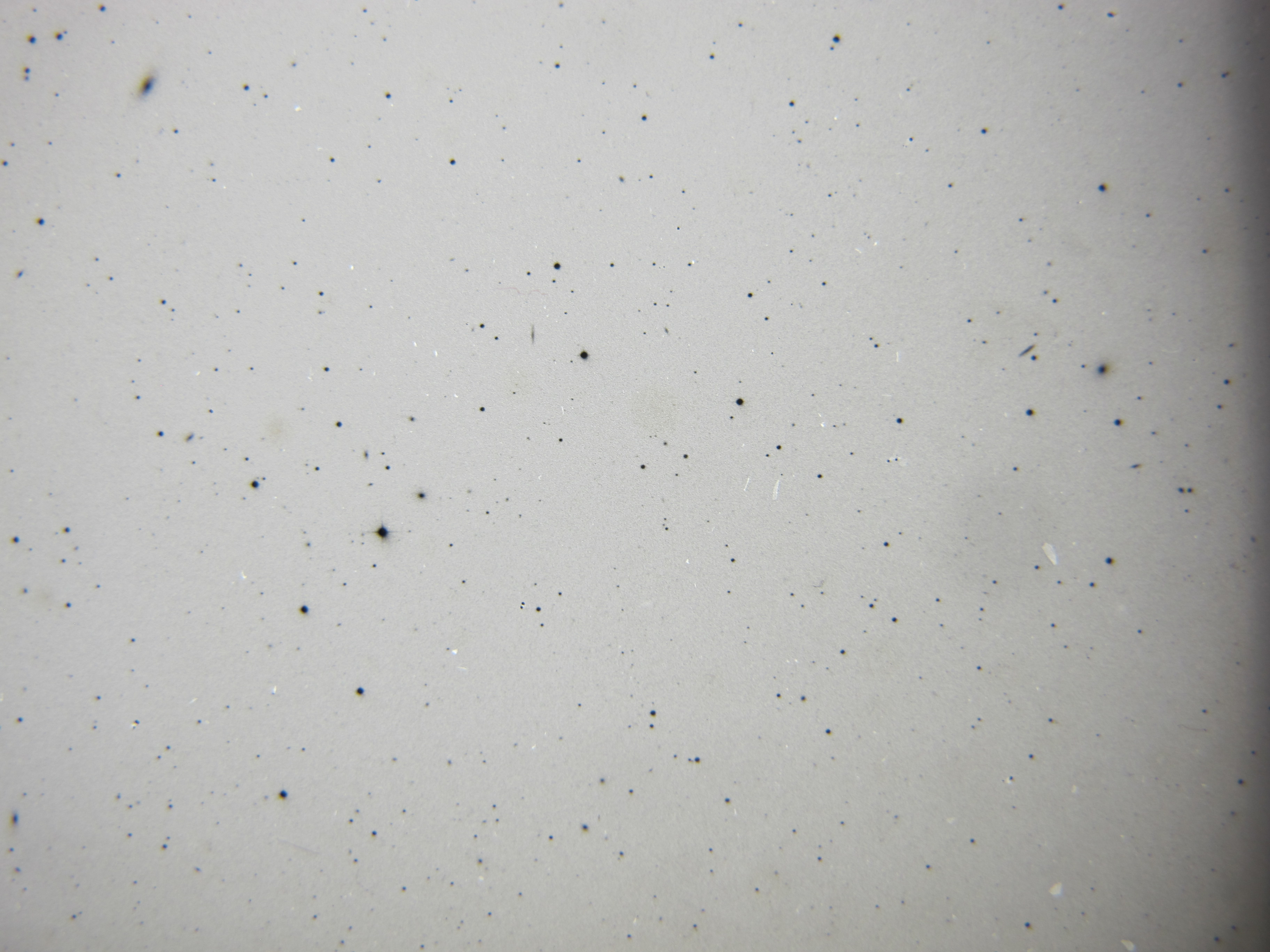}
    \caption{Highest-resolution microscopy image of the XE325 plate field. This image provides the clearest view of the morphological differences between the transient candidates (sharp, compact) and typical stellar images (diffuse edges, some elongation). All nine objects from the original detection are confirmed visible on the physical plate, ruling out a digitization-artifact origin. The sharp-edged morphology is consistent with both very short-duration astrophysical flashes and with emulsion defects, particularly given the plate-corner location.}
    \label{fig:microscopy_9757}
\end{figure}

The inspection was conducted using a $25\times$ magnifier and a hand-held digital camera. All nine objects identified in the field were visible on the original plate (Figures~\ref{fig:microscopy_9755}--\ref{fig:microscopy_9757}). Microscopy reveals that these sources appear sharper, with more distinct edges, than the stellar images, which have diffuse edges. The transient candidates also show no elongation, whereas stellar images in the field exhibit some elongation. This morphology is qualitatively consistent with expectations for very short-duration flashes integrated over a long exposure, which would produce more compact point spread functions (PSFs) than stars.

However, the field of interest is located in the corner of the physical plate. The microscopist's observational report described the features as ``not stellar in appearance'' with ``sharp edges,'' which, in the absence of knowledge about transient phenomena, could be attributed to emulsion defects, which are more common in plate corners\footnote{From the microscopist's letter of 14 December 2025: ``[X] \textit{and I both looked at the 9 objects on your field, being directed to them by Mr.~Hambly. First of all, all 9 objects were visible on the original plate. But in my opinion, they were not stellar in appearance. The edges were sharp, unlike the stellar images, which had diffuse edges. There was no elongation, as was present on the stellar images. I'd like to add that this field of interest is also in the corner of the physical plate. In my experience, it is reasonable to attribute these anomalies to emulsion defects.}''}. This highlights the central interpretative ambiguity: the same morphological characteristics --- sharp, compact, non-elongated --- that are consistent with sub-second flashes are also consistent with certain types of plate defect. We return to this point in Section~\ref{sec:discussion}, where we stress the need to separate the (factual) morphological observation from its (model-dependent) interpretation.

\subsection{Case 2: Inspection of copy plate at Harvard}

For the triple-transient event (XE186), we inspected a copy plate held in the Harvard College Observatory collection. The copy plate reproduces the transient features seen in the original Palomar data, providing an independent confirmation that the signals are present in the photographic material and were not introduced during digitization.

\begin{figure}
    \centering
    \begin{subfigure}[b]{0.48\linewidth}
        \centering
        \includegraphics[width=\linewidth]{micro.jpg}
        \caption{Harvard copy plate (microscopy).}
        \label{fig:compare_case2_micro}
    \end{subfigure}
    \hfill
    \begin{subfigure}[b]{0.48\linewidth}
        \centering
        \includegraphics[width=\linewidth]{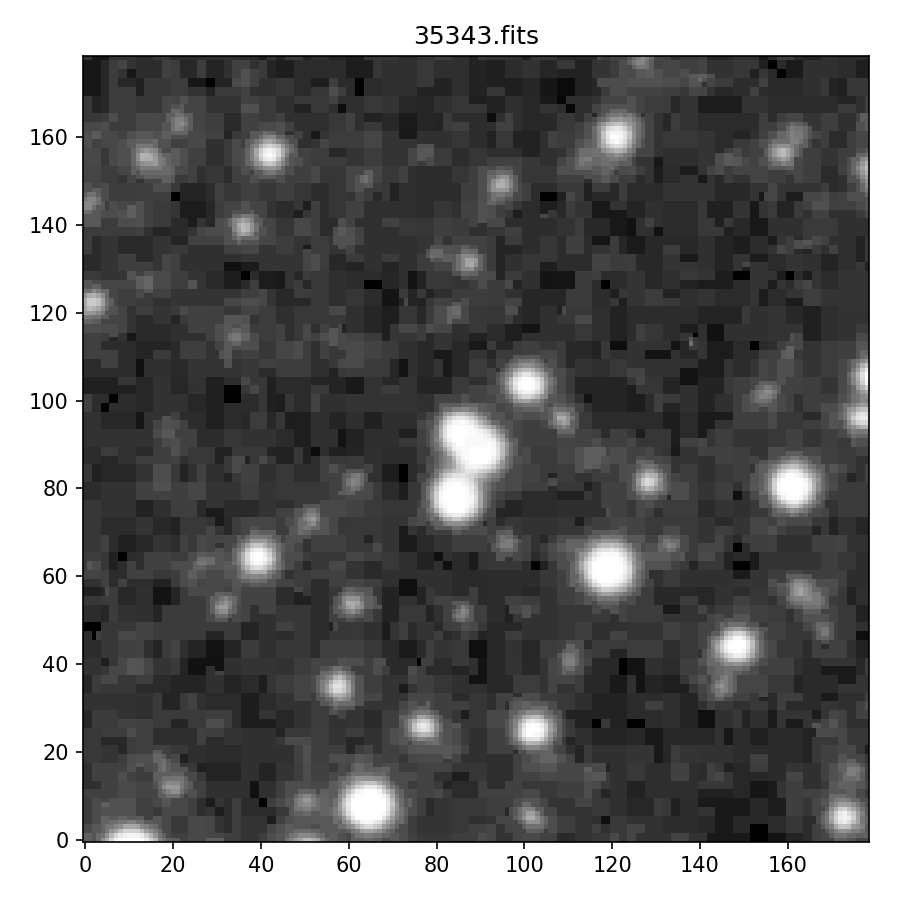}
        \caption{Original survey digitization.}
        \label{fig:compare_case2_digi}
    \end{subfigure}
    \caption{Side-by-side comparison of the triple-transient field (XE186) between the Harvard copy-plate microscopy (left) and the published original-plate digitization (right), registered using the brighter field stars. The comparison is the actual discriminant for the copy-plate-artefact hypothesis of \citet{Hambly2024}: a candidate that is present in the original digitization at the catalogued position is intrinsic to the original exposure, whereas a feature appearing only in the copy material would indicate a copying artefact. The three catalogued sources of \citet{Solano2024} are recovered in both representations at their tabulated positions.}
    \label{fig:compare_case2}
\end{figure}

The copy plate is, however, exactly the material for which \citet{Hambly2024} argued that star-like ``transients'' can be introduced by the copying procedure used to disseminate the survey atlas before large-scale digitization. The copying process can add smoothing and contrast changes, and it can imprint its own grain. The relevant test is therefore not whether the field looks grainy under the microscope --- all emulsion does --- but whether the transient candidate is also present, at the same position, in the \emph{original} survey digitization. We address this by registering the copy-plate microscopy field to the SuperCOSMOS and STScI scans of \citet{Solano2024} (Figure~\ref{fig:compare_case2}).

\begin{figure}
    \centering
    \includegraphics[width=0.45\textwidth]{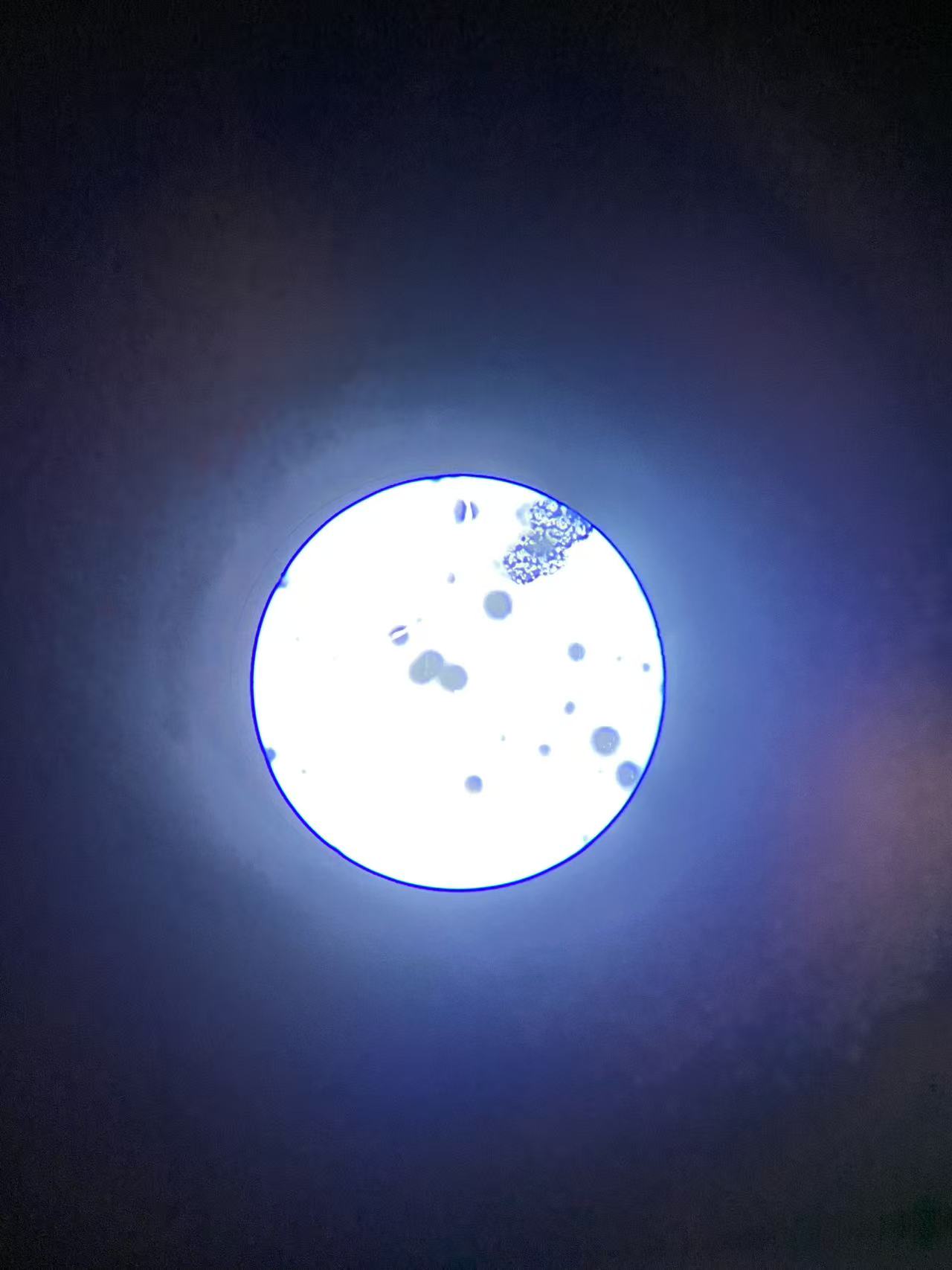}
    \caption{Microscopy image of the photographic-plate emulsion obtained from the Harvard copy plate. The bright circular region is the microscope field of view. Several compact, approximately round features are visible, consistent with stellar PSFs, together with the diffuse halo and fine grain characteristic of the copy-plate emulsion. The triple-transient candidate is present and appears as a compact, localized feature, confirming that it is not introduced by digitization. Because copy-plate grain and copying artefacts can themselves mimic compact sources \citep{Hambly2024}, the presence of the feature here establishes its reality \emph{at the plate level} but does not by itself distinguish a short-duration flash from an emulsion defect; that requires comparison with the original digitization (Figure~\ref{fig:compare_case2}) and, ultimately, population-level statistics.}
    \label{fig:microscopy}
\end{figure}

\begin{figure}
    \centering
    \begin{subfigure}[b]{0.48\linewidth}
        \centering
        \includegraphics[width=\linewidth]{IMG_9757.JPG}
        \caption{Caltech original plate (magnifying glass).}
        \label{fig:compare_case1_micro}
    \end{subfigure}
    \hfill
    \begin{subfigure}[b]{0.48\linewidth}
        \centering
        \includegraphics[width=\linewidth]{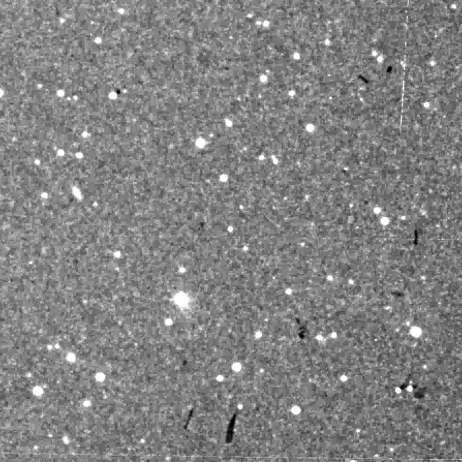}
        \caption{Original survey digitization.}
        \label{fig:compare_case1_digi}
    \end{subfigure}
    \caption{Side-by-side comparison of the nine-transient field (XE325) between the Caltech original-plate (left) and the published digitization \citealt{Villarroel2021} (right), registered using the brighter field stars. All nine catalogued sources are located in the magnifying glass frame at their published positions, confirming that the features are present on the physical emulsion rather than being digitization artefacts. As for Case~2, this registration establishes co-location and reality at the plate level but does not classify individual sources as flash versus defect.}
    \label{fig:compare_case1}
\end{figure}

\subsection{Cross-matching with other archives}
\label{sec:crossmatch}

A natural test of the physical reality of a short-duration optical flash is to ask whether the same event was recorded, at the same time, from a second site. Two independent, simultaneous detections would confirm a celestial (or near-Earth) origin and, for nearby objects, would allow a parallax distance; a detection from only one site would be consistent with a plate defect or a very local phenomenon.

We searched the StarGlass and DASCH plate metadata for exposures covering the XE325 and XE186 fields on the relevant nights (12 April 1950 and 19 July 1952). In practice the requirements for a usable confirmation are severe and, for these two events, are not met by existing material:

\begin{itemize}
    \item \textit{Simultaneity.} The candidates are consistent with emission lasting $\lesssim$ a few seconds within a $\sim 50$-min exposure. A second-site plate must overlap the same few-second window to record the same flash; mere coverage of the field on the same night is insufficient. Patrol programmes of the era were not synchronised to this precision.
    \item \textit{Depth.} The sources are $R \sim 16$ at peak \citep{Solano2024} but the discovery plates reach $r \sim 20$; a confirming plate must reach comparable depth to rule out, rather than simply fail to detect, a counterpart. Contemporaneous wide-field patrol plates (e.g.\ the Sonneberg, Bamberg, and Harvard patrol series) typically have substantially brighter limiting magnitudes and would not register sources at these levels even if they were present.
    \item \textit{Astrometric and proper-motion considerations.} If the sources are nearby (as the causality argument of \citealt{Solano2024} allows for XE186), parallax and proper motion over the relevant baselines must be folded into the search radius; this further enlarges the required search area and the astrometric tolerance.
\end{itemize}

Given these constraints, our archival search does not yield a usable independent confirmation for either event. We stress that this is a statement about the \emph{feasibility of the test}, not a null result: because no simultaneous second-site plate of comparable depth exists for these fields, the multi-observatory test cannot presently be performed, and its non-availability should not be read as evidence against the reality of the events. The approach is, however, directly actionable for future, purpose-built searches (Section~\ref{sec:multiobs}).

\section{Results}

Our analysis yields three main findings.

First, the transient candidates in both cases exhibit point-like morphologies that are broadly consistent with stellar PSFs but appear marginally sharper. This is qualitatively consistent with expectations for sub-second to few-second flashes occurring during exposures of $\sim 50$~min, as previously suggested.

Second, the features are reproducible across different representations of the data. In the nine-transient case the signals are visible on the original glass plate; in the triple-transient case they are present both in the independent copy plate and, crucially, in the original survey digitization at the catalogued positions (Figures~\ref{fig:compare_case2} and \ref{fig:compare_case1}). This reduces the likelihood that the signals are purely digital or scanning artefacts and, for Case~2, disfavours the copy-plate-artefact hypothesis of \citet{Hambly2024} for these specific sources, since the features are not confined to the copy material.

Third, despite these suggestive properties, microscopy alone does not provide definitive evidence to distinguish true transient events from plate defects. It is well established that the grain structure of photographic emulsions can produce compact, point-like features, and that such features have been proposed as the origin of the copy-plate transients \citep{Hambly2024}; our own material confirms that the candidate fields are grainy at high magnification. What our material does \emph{not} allow is the per-source classification of individual catalogued features as ``defects mimicking astrophysical sources'': registering the microscopy to the published digitization (Section~\ref{sec:registration}) establishes that the catalogued sources are present on the physical emulsion at their expected positions, but it does not, on morphology alone, assign any given source to the flash or the defect category. In the XE325 case the microscopist noted sharp edges and a lack of elongation --- characteristics that differ from typical stellar images but that could be read as emulsion defects, particularly given the plate-corner location. That reading reflects a natural bias: in the absence of the transient hypothesis, the default comparison is with steadily shining stars, which favours a defect interpretation over sub-second emission.


\section{Discussion}
\label{sec:discussion}

The morphology of the observed transients is consistent with expectations for short-duration flashes. If an object emits light for a very brief interval during a long exposure, its image will not streak but will instead produce a compact PSF. This interpretation is supported by the slightly sharper appearance of the transient candidates relative to background stars.

However, another plausible explanation for the morphologies alone is that these features arise from intrinsic plate defects or emulsion irregularities. The difficulty is that both scenarios produce qualitatively similar signatures at the level of individual events. The XE325 inspection illustrates this: the microscopist's observation that the features were ``sharp'' and ``round'' is factual, but the further statement that they are ``reasonably attributed to emulsion defects'' is an interpretation made by comparison with steady stellar sources, without considering transient phenomena. Observation and interpretation must therefore be kept distinct.

A key implication is that microscopy is insufficient, by itself, to resolve this ambiguity. Microscopy can confirm that a feature is present on the original plate (ruling out digitization artefacts), can identify obvious holes or gross defects, and can document morphology; but it cannot determine whether a sharp, compact feature represents a brief flash or a localized emulsion defect. For individual events, the more powerful approach is that of \citet{Busko2026}, who examined the optical aberrations of transient sources and showed that, despite their narrower FWHM, the transients exhibit the same aberrations as ordinary astronomical sources, and therefore originate from light that passed through the telescope optics. At the population level, statistical approaches --- spatial distribution, recurrence rate, and correlation with observing conditions --- are required.

\subsection{Prospects for multi-observatory confirmation}
\label{sec:multiobs}

Simultaneous observations from two or more sites would provide a clean, model-independent test: a flash recorded at both sites is celestial or near-Earth in origin, and its parallax constrains the distance, which is especially diagnostic for near-Earth objects. As set out in Section~\ref{sec:crossmatch}, however, this test is not feasible for XE325 or XE186 with existing plates, because no contemporaneous second-site exposure of comparable depth and sub-exposure timing exists for these fields. The value of the multi-observatory approach is therefore prospective rather than retrospective for the historical sample.

For future searches the method is straightforward to implement: two (or more) stations separated by a baseline of order $10^2$--$10^3$~km, imaging the same field synchronously with deep, short exposures, would either record a transient at both sites --- yielding an immediate parallax --- or at only one, flagging a local or instrumental origin. Such a configuration converts the ambiguous single-plate detections that motivate this paper into decisive measurements. We note that a comparable logic underlies the Earth-shadow filtering already applied to modern survey data \citep{VillarroelPASP,Doherty2026}.

\section{Conclusion}

We have presented a microscopy-based assessment of two representative multiple-transient candidates identified on historical photographic plates. The observed features are consistent with short-duration transient events and are reproducible across the original plate, the copy plate, and the published digitization, strengthening the case that they are not artifacts introduced during digitization. At the same time, our analysis demonstrates a fundamental limitation of microscopy: the appearance of an individual point-like feature cannot by itself distinguish a genuine transient flash from an intrinsic plate defect. Also, unfortunately, the multi-observatory parallax test cannot be applied to these particular events because no contemporaneous plates exist from an independent site.

These results therefore sharpen, rather than resolve, the central question. They highlight the fundamental limitations of single-event analyses and clarify what microscopy can, and cannot, establish for individual candidates. Resolving the nature of these events will require population-level studies of large archival datasets and, for future events, purpose-built multi-station imaging. Combining large archival datasets with modern time-domain surveys may provide the statistical power needed to resolve the nature of these intriguing events.

\section*{Acknowledgements}

We thank the Caltech team for providing access to the original plate imagery and for their assistance with the microscopy analysis, and the Harvard College Observatory plate stacks team for facilitating access to the copy plate and archival material. We thank Avi Loeb for helpful discussions and scientific input. JY acknowledges support from private donors.

\section*{Data Availability}

The microscopy images presented here are available from the corresponding author on reasonable request. The POSS-I digitizations are publicly available through the STScI and SuperCOSMOS archives, and the archival plate metadata through the StarGlass and DASCH services.

\bibliographystyle{mnras}
\bibliography{ss-seti_bv}


\bsp	
\label{lastpage}
\end{document}